\documentclass{article}
\usepackage{spconf,amsmath,graphicx,hyperref}
\usepackage{amsmath,amssymb,amsfonts}
\usepackage{algorithmic}
\usepackage{graphicx}
\usepackage{subcaption}
\usepackage{tabularray}
\UseTblrLibrary{amsmath,booktabs,counter,diagbox,nameref,siunitx,varwidth,zref}
\usepackage{float}

\usepackage{textcomp}
\usepackage{xcolor}
\usepackage{booktabs}
\usepackage{cleveref}
\usepackage{placeins}
\usepackage{tikz}
\usepackage{pgfplots}
\pgfplotsset{compat=1.18}
\usepackage{pgfplotstable}
\usepackage{amsmath}
\usepackage{xcolor}
\usepgfplotslibrary{groupplots}
\usetikzlibrary{positioning}
\usepackage{pifont}
\usepackage{url}
\usetikzlibrary{plotmarks}

\definecolor{myred}{HTML}{E76F51}
\definecolor{myblue}{HTML}{376996}
\definecolor{mygreen}{HTML}{2A9D8F}

\pgfplotsset{
  mylegendstyle/.style={
    legend image code/.code={
      \draw[##1, fill=##1, fill opacity=0.5] (0cm,-0.1cm) rectangle (0.3cm,0.25cm);
    }
  }
}

\title{HyperCool: Reducing Encoding Cost \\ in Overfitted Codecs with Hypernetworks}

\name{Pep Borrell-Tatché$^1$ \qquad Till Aczel$^1$ \qquad Théo Ladune$^2$ \qquad Roger Wattenhofer$^1$}
\address{$^1$ETH Zürich \qquad $^2$Orange Research}
\begin{document}
\ninept
\maketitle
\begin{abstract}
Overfitted image codecs like Cool-chic achieve strong compression by tailoring lightweight models to individual images, but their encoding is slow and computationally expensive.
To accelerate encoding, Non-Overfitted (N-O) Cool-chic replaces the per-image optimization with a learned inference model, trading compression performance for encoding speed.  
We introduce HyperCool, a hypernetwork architecture that mitigates this trade-off.
Building upon the N-O Cool-chic framework, HyperCool generates content-adaptive parameters for a Cool-chic decoder in a single forward pass, tailoring the decoder to the input image without requiring per-image fine-tuning.
Our method achieves a 4.9\% rate reduction over N-O Cool-chic with minimal computational overhead.
Furthermore, the output of our hypernetwork provides a strong initialization for further optimization, reducing the number of steps needed to approach fully overfitted model performance.
With fine-tuning, HEVC-level compression is achieved with 60.4\% of the encoding cost of the fully overfitted Cool-chic.  
This work proposes a practical method to accelerate encoding in overfitted image codecs, improving their viability in scenarios with tight compute budgets.
\end{abstract}
\begin{keywords}
Image compression, learned compression, lightweight models, per-image overfitting.
\end{keywords}
\section{Introduction}
% Image compression is a critical component of modern visual data processing.
Learned image compression methods can outperform traditional codecs in rate-distortion (RD) performance, particularly at low bitrates \cite{liu_learned_2023, jiang_mlic_2025}.
These methods train neural networks end-to-end to optimize RD metrics, but often impose substantial computational demands.
To address the decoding cost,  Cool-chic \cite{ladune_Cool-chic_2023} and the C3 framework \cite{kim_c3_2024} introduce a novel approach: instead of relying on large, fixed, pre-trained models, they overfit lightweight neural networks to individual images and transmit the network parameters as the compressed representation.
This per-image overfitting yields competitive compression with minimal decompression cost, offering a compelling alternative to autoencoder and diffusion-based schemes, which remain compute-intensive, particularly at decode time.

\pgfdeclareplotmark{sharp star}{
  \pgfpathmoveto{\pgfpointpolar{90}{2pt}} % Start at top point rotated -90°, so 0°
  \foreach \i in {1,...,5} {
    \pgfpathlineto{\pgfpointpolar{90 + \i*144 - 72}{1pt}} % inner points
    \pgfpathlineto{\pgfpointpolar{90 + \i*144}{2pt}}     % outer points
  }
  \pgfpathclose
  \pgfusepathqfillstroke
}

\begin{figure}[!t]
  \centering
  \begin{tikzpicture}
    \begin{semilogxaxis}[
        width=\linewidth,
        height=7cm,
        xlabel={Encoding complexity [MAC / pixel] $\downarrow$},
        ylabel={BD-rate vs. HEVC (HM 16.20) [\%] $\downarrow$ },
        grid=major,
        minor y tick num=1,
        % yminorgrids=true,
        ymin=-10, ymax=30,
        xmin=50000, xmax=50000000,
        ytick distance={10},
        legend columns=3,
        legend style={
        font=\footnotesize\sffamily,
        at={(0.45,-0.25)},
        anchor=north,
        draw=black,
        % column sep=1.5em,
        },
        every axis/.append style={
            line width=0.9pt,
            tick label style={font=\footnotesize},
            label style={font=\footnotesize},
            title style={font=\footnotesize, yshift=-1ex},
        },
        legend cell align={left},
    ]

    % Scatter: Initial points
    % \addplot[only marks, mark=*, mark size=2pt, color=myblue] coordinates {
    %    (22.208348, 36.233905)
    % };
    % \addlegendentry{N-O Cool-chic}

    % \addplot[only marks, mark=triangle*, mark size=4pt, color=myred] coordinates {
    %    (27.636784, 28.284959)
    % };
    % \addlegendentry{\textbf{Hypernet (ours)}}

    % Line: Hypernet finetuning
    \addplot[
        mygreen,
        ultra thick,
        mark=*,
        mark size=2pt,
        nodes near coords, % Place nodes near each coordinate
        point meta=explicit symbolic, % The meta data used in the nodes is not explicitly provided and not numeric
        every node near coord/.style={
            anchor=north,
            font=\footnotesize,
            yshift=-2pt,
        }
    % meta index = 3 --> look for the node label (the number of training iteration in the 4th column
    ] table [x=mac_pixel, y=bdrate, meta index=3] {data/main_figure/hypernet.tsv};
    \addlegendentry{\textbf{HyperCool (ours)}}

    % Line: N-O Cool-chic finetuning
    \addplot[
        myblue,
        dashed,
        mark=triangle*,
        mark size=2pt,
        mark options={solid},
        nodes near coords, % Place nodes near each coordinate
        point meta=explicit symbolic, % The meta data used in the nodes is not explicitly provided and not numeric
        every node near coord/.style={
            anchor=south,
            font=\footnotesize,
            yshift=2pt,
        }
        % meta index = 3 --> look for the node label (the number of training iteration in the 4th column
        ] table [x=mac_pixel, y=bdrate, meta index=3] {data/main_figure/no-Cool-chic.tsv};
    \addlegendentry{N-O Cool-chic}

    \addplot[
        myred,
        dashed,
        mark=sharp star,
        mark size=2pt,
        mark options={solid},
        nodes near coords, % Place nodes near each coordinate
        point meta=explicit symbolic, % The meta data used in the nodes is not explicitly provided and not numeric
        every node near coord/.style={
            anchor=west,
            font=\footnotesize,
            xshift=2pt
        }
    % meta index = 0 --> look for the node label (the number of training iteration in the 1st column
    ] table [x=mac_pixel, y=bdrate, meta index=0] {data/main_figure/coolchic.tsv};
    \addlegendentry{Cool-chic 4.0}

    % Horizontal line at BD-rate = 0 (HEVC reference)
    \addplot[
        black,
        dashed,
        domain=50000:50000000,
    ]
    {0}
    node[pos=0.05, anchor=north west, font=\footnotesize\bfseries, text=black] {HEVC (HM 16.20)};

    \end{semilogxaxis}
  \end{tikzpicture}
  \caption{
    BD-rate against encoding complexity when fine-tuning from different initializations on the CLIC2020 dataset. Numbers next to data points indicate optimization steps.
  }
  \label{fig:bd-vs-flops}
\end{figure}
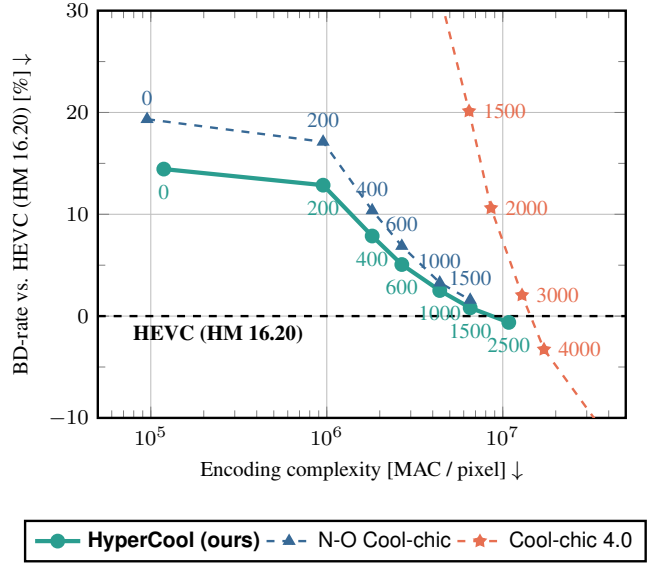

\begin{figure*}[!t]
  \centering
  \includegraphics[width=0.95\linewidth]{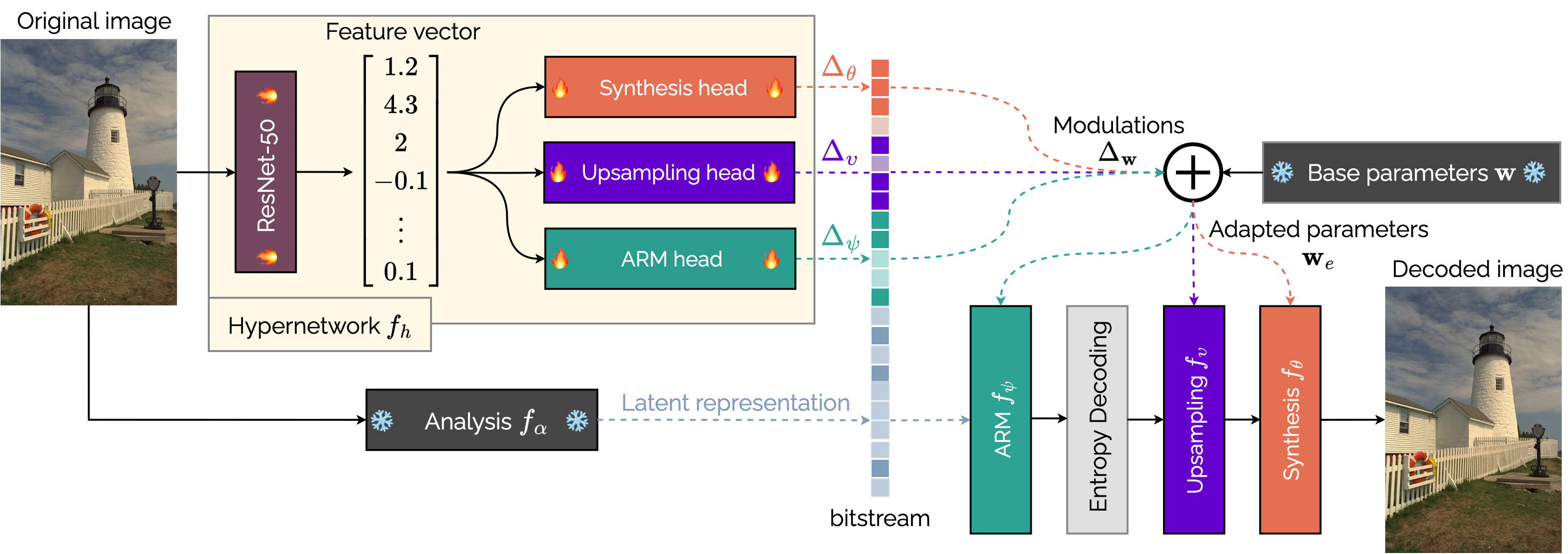}
  \caption{Architecture of the proposed HyperCool. The hypernetwork takes an input image and produces weight modulations for the synthesis, upsampling, and ARM composing a Cool-chic decoder. Only the weight modulations are transmitted.}
  \label{f:hypernet-arch}
\end{figure*}

Despite its fast decoding, Cool-chic suffers from slow encoding, requiring iterative optimization of both weights and latents from scratch per image.
To address this, Blard et al. propose Non-Overfitted (N-O) Cool-chic \cite{blard_overfitted_2024}, which replaces per-image optimization with an analysis transform and a universal decoder that produces latents directly, without iterative rate-distortion optimization.
This yields a substantial encoding speed-up and maintains Cool-chic's low decoding complexity.
However, it also degrades compression efficiency, incurring a 56.5\% rate increase on the CLIC2020 dataset.

This work aims to recover the compression efficiency lost in N-O Cool-chic while retaining its fast encoding and low decoding cost.
We introduce \emph{HyperCool}, a new variant of Cool-chic that restores image-dependent information in the decoder by employing a hypernetwork to predict decoder weights conditioned on the input image.
HyperCool improves compression performance over N-O Cool-chic while retaining its fast encoding and maintaining the same low decoding cost.
On the CLIC2020 dataset, it achieves a 4.9\% BD-rate reduction compared to N-O Cool-chic, narrowing the gap to fully overfitted methods.

In addition to providing fast and adaptive compression, HyperCool supports optional fine-tuning of the predicted decoder on a single image, effectively using it as a warm start for full Cool-chic overfitting.
This hybrid strategy reaches HEVC-level compression while requiring only 60.4\% of the original Cool-chic encoding cost and preserving its decoding efficiency.
We also provide a detailed analysis of the trade-offs between hypernetwork inference, optional per-image fine-tuning, and the resulting rate–distortion performance.

\section{Related work and background}\label{s:relatedw}

\subsection{Learned Image Compression}
Learned image compression typically uses autoencoder-based architectures \cite{balle_end--end_2017, balle_variational_2018}.
An encoder maps the image $\mathbf{x}$ to latents $\mathbf{y}$, which are quantized to $\hat{\mathbf{y}}$ and entropy-coded.
A decoder reconstructs the image from $\hat{\mathbf{y}}$.
These models are trained end-to-end with a rate-distortion loss balancing reconstruction quality and bitrate:
\begin{equation}
  \mathcal{L} = R(\hat{\mathbf{y}}) + \lambda D(\mathbf{x}, \hat{\mathbf{x}})
  \label{eq:rd-loss}
\end{equation}
where $D$ is a distortion metric (e.g., MSE), $R$ estimates the bitrate, and $\lambda$ controls the trade-off.

% Early works \cite{balle_end--end_2017} approximated quantization with uniform noise during training to enable gradient descent.
% Hyperprior models \cite{balle_variational_2018} introduced side information to improve entropy estimation.
% Autoregressive models \cite{minnen_joint_2018} and more expressive priors with attention \cite{cheng_learned_2020} further enhanced rate-distortion performance.
% Perception-driven losses \cite{mentzer_high-fidelity_2020, he_po-elic_2022} improved visual quality, especially at low bitrates.
\subsection{Overfitted Codecs}
Overfitted codecs train a dedicated model per image.
COIN \cite{dupont_coin_2021} encodes each image as a fully connected network mapping coordinates to RGB values.
COIN++ \cite{dupont_coin_2022} introduces a meta-learned base network shared across images and small per-image modulations, which are quantized and entropy-coded.

Cool-chic \cite{ladune_Cool-chic_2023} extends these ideas by: 
(1) Representing images with hierarchical latent grids $\hat{\mathbf{y}} = {\hat{y}_1, \ldots, \hat{y}_N}$ capturing multi-scale structure.
(2) Using a small synthesis network $f_\theta$ to reconstruct images from upsampled latents.
(3) Compressing latents with an image-specific autoregressive entropy model $f_\psi$ conditioned on causal context.

Encoding in Cool-chic requires overfitting $\{\hat{\mathbf{y}}, \theta, \psi\}$ per image by minimizing a rate-distortion loss:
\begin{equation}
  \mathcal{L} = \mathbb{E}_{\mathbf{x}}\left[
    \lambda D(\mathbf{x}, f_\theta(\hat{\mathbf{y}})) - \log p_\psi(\hat{\mathbf{y}})
  \right],
\end{equation}
where $p_\psi$ is modeled autoregressively:
\begin{equation}
  p_\psi(\hat{\mathbf{y}}) = \prod_{i,j,k} p_\psi(\hat{y}_{ijk} | \mathbf{c}_{ijk}).
\end{equation}
Cool-chic offers strong compression with a lightweight decoder but requires thousands of gradient steps per image, resulting in slow encoding.

Subsequent works improved Cool-chic via architectural refinements, improved quantization, and training strategies \cite{kim_c3_2024,leguay_low-complexity_2023,philippe_upsampling_2024}. 
The Cool-chic open-source implementation \cite{coolchic} integrates most improvements and serves as our starting point.

\subsection{Reducing Encoding Complexity}
Non-Overfitted (N-O) Cool-chic \cite{blard_overfitted_2024} speeds up encoding by removing per-image optimization and learning:
(1) An analysis transform $f_\alpha$ that maps images to latents in a single forward pass.
(2) A universal upsampling, synthesis network, and entropy model.
The model is trained end-to-end by minimizing:
\begin{equation}
  \min_{\alpha, \theta, \psi} \mathbb{E}_\mathbf{x} \left[
    \lambda D(\mathbf{x}, f_\theta(\text{Ups}(f_\alpha(\mathbf{x})))) - \log p_\psi(f_\alpha(\mathbf{x}))
  \right].
\end{equation}
N-O Cool-chic enables fast encoding but loses some compression efficiency relative to fully optimized Cool-chic.

Metalearning methods like MLIIC \cite{zhang_mliic_2025} use meta-learned initializations to speed up adaptation and boost compression, but the code is unreleased and the results unverified.

\begin{figure*}[!tb]
  \centering
    \begin{subfigure}[b]{0.49\linewidth}
        \includegraphics[width=\linewidth]{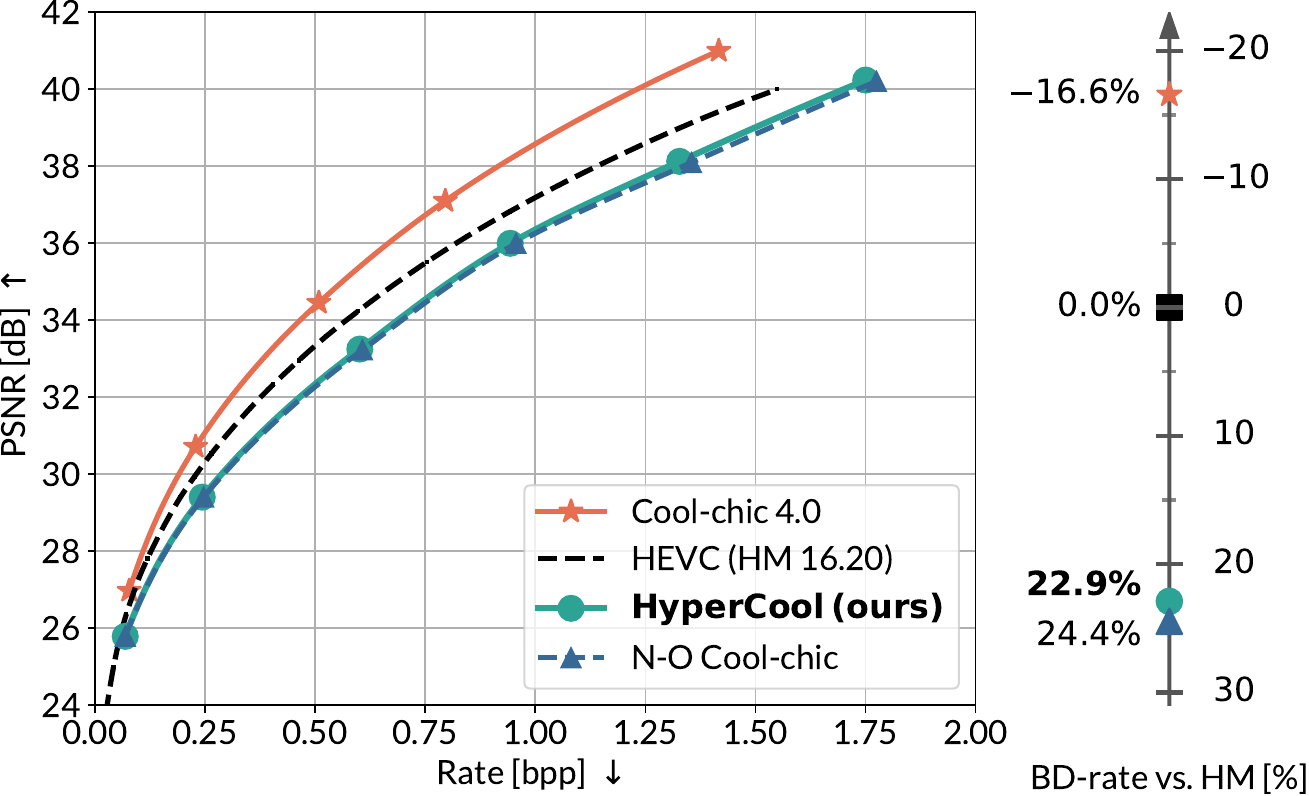}
        \caption{Kodak dataset}
    \end{subfigure}
    \hfill
    \begin{subfigure}[b]{0.49\linewidth}
        \includegraphics[width=\linewidth]{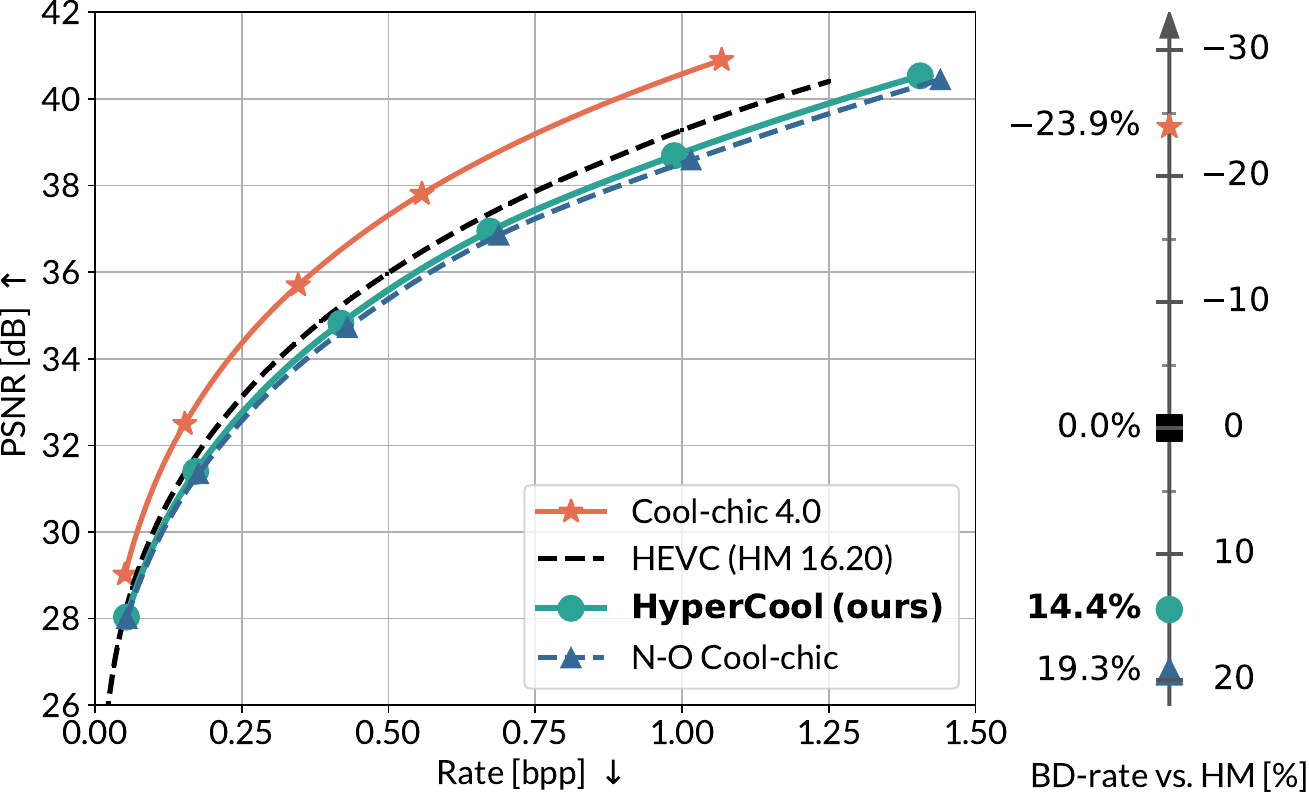}
        \caption{CLIC2020 pro validation dataset}
    \end{subfigure}
  \caption{
    HyperCool rate-distortion performance. Results are averaged across the whole test dataset.
  }
  \label{fig:avg-rds}
\end{figure*}

\section{Method}\label{s:method}

We propose a hypernetwork-based method that merges N-O Cool-chic’s efficiency with the adaptability of overfitted decoders, reducing encoding time while boosting compression.
Figure \ref{f:hypernet-arch} illustrates the encoding and decoding process.
% \newline

Starting from a pretrained N-O Cool-chic base model with decoder parameters
% \footnote{Parameters include weights and biases; some sections of the Cool-chic architecture include biases while others do not.} 
$\mathbf{w}$ and an analysis transform $f_\alpha$ mapping images to latent grids $\hat{\mathbf{y}}$, we train a hypernetwork $f_h$ to produce image-conditioned \emph{modulation parameters} $\Delta_\mathbf{w}$:
\begin{equation}
    \Delta_\mathbf{w} = f_h(\mathbf{x}).
\end{equation}
As shown in Figure \ref{f:hypernet-arch}, the hypernetwork $f_h$ consists of two parts: 
A pretrained \texttt{ResNet-50} backbone extracts features from $\mathbf{x}$.
This is followed by separate MLP heads generating modulations for the upsampling, synthesis, and autoregressive entropy model.
% \newline

The modulation $\Delta_\mathbf{w}$ is transmitted alongside the latent representation $\hat{\mathbf{y}}$.
Modulations are encoded like standard Cool-chic neural network parameters: first quantized, then entropy-coded using Exp-Golomb coding.
To decode the image, the image-adapted parameters $\mathbf{w}_e$ are constructed by adding the base decoder parameters $\mathbf{w}$ and the modulation $\Delta\mathbf{w}$:
\begin{equation}
  \mathbf{w}_e = \mathbf{w} + \Delta_\mathbf{w}.
  \label{e:image-adapted-param}
\end{equation}
These image-adapted parameters are then used to compute the decoded image from the latent representation.
% \newline

At inference, the hypernetwork predicts modulation parameters in a forward pass.
These modulations adapt the decoder to the image, improving compression.
However, transmitting the modulations introduces a small rate overhead.
At the encoder side, a test verifies if the modulations actually improve overall performance.
If not, they are discarded.
This ensures our method never performs worse than the base N-O Cool-chic and often improves upon it.

\section{Results}

\subsection{Training}

HyperCool is trained on 500{,}000 samples from the OpenImages dataset \cite{kuznetsova_open_2020}. 
Following standard practice, training uses random $256 \times 256$ patches extracted from the training dataset.
The hypernetwork is learned on top of pretrained N-O Cool-chic models\footnote{We thank \textbf{Théophile Blard} for training these models.} using our method.
One optimization step consists of the encoding and decoding described in Section \ref{s:method} and depicted in Fig. \ref{f:hypernet-arch}.
% \newline

Only the hypernetwork parameters $h$ are trained \textit{i.e.}, the Res-Net50 backbone and the different MLP heads. All the N-O Cool-chic parameters remain fixed, including the base decoder parameters $\mathbf{w}$ and analysis transform $f_\alpha$.
Since the latent is not optimized, latent quantization remains non-differentiable, simplifying training.
The training loss is the standard rate-distortion objective:
\begin{equation}
  h = \arg\min \mathbb{E}_{\mathbf{x}}\left[
    \lambda D(\mathbf{x}, \hat{\mathbf{x}}) + R(\hat{\mathbf{y}})
  \right],
\end{equation}
where the decoded image $\hat{\mathbf{x}}$ and latent rate $R(\hat{\mathbf{y}})$ are obtained using the image-adapted parameters $\mathbf{w}_e$.
Note that during training, the rate term only accounts for the latent representation’s bitrate (via the adapted ARM), excluding the modulation parameters’ rate.

\subsection{Compression and encoding complexity trade-off}

We evaluated our methods on the Kodak \cite{noauthor_kodak_nodate} and CLIC2020  professional validation \cite{noauthor_workshop_2020} datasets.
Kodak contains 24 images at $768 \times 512$ resolution, while CLIC2020 includes 41 images ranging from $512 \times 384$ to $2048 \times 1370$.
% \newline

% We use the Bj{\o}ntegaard delta rate (BD-rate) to compare compression methods.
% BD-rate measures the average difference in bitrate between two rate-distortion curves over a shared quality range.
% It reflects the relative bitrate needed to match the quality of a reference method.

\Cref{fig:avg-rds} shows the rate-distortion performance of HyperCool compared to the N-O Cool-chic baseline and the original overfitted Cool-chic 4.0. 
Our method improves compression over N-O Cool-chic on both datasets. 
Gains are more pronounced at higher bitrates and on larger images, such as those in CLIC2020.
% \newline

\begin{table}[!ht]
    \centering
    \caption{
      Encoding complexity and BD-rate against HEVC of the proposed HyperCool compared to N-O Cool-chic.
    }
    \small
    \begin{tblr}{
        colspec={Q[c] Q[c] Q[c] Q[c] Q[c] Q[c] Q[c]},
        row{1} = {darkgray!5},
        row{2} = {darkgray!5},
        cell{1}{1}={c=1,r=2}{font=\bfseries},
        cell{1}{2}={c=3,r=1}{font=\bfseries},
        cell{1}{5}={c=2,r=1}{font=\bfseries},
        colsep=3.5pt,
        rowsep=2pt,
        vline{2}={1-Z}{solid, fg=darkgray,1pt},
        vline{5}={1-Z}{solid, fg=darkgray,1pt},
    }
        Method          & Complexity [kMAC / pix] $\downarrow$  &  & & BD-rate [\%] $\downarrow$        \\
                        & Analysis                  & Hypernet & Total & Kodak    & CLIC20 \\
        \cmidrule[1pt,darkgray]{1-Z}
        N-O Cool-chic   & 99                      & /        & 99  & 24.4 & 19.3 \\      
        HyperCool         & 99                      & 24     & 123 & 22.9 & 14.4 \\      
        % NEW RESULTS: clic: 14.4, kodak: 22.87
        \cmidrule[1pt,darkgray]{1-Z}
        Cool-chic fast  & / & / & 64~000         & -11.8  & -16.9 \\      
        Cool-chic slow  & / & / & 450~000        & -16.6    & -23.9 \\      
    \end{tblr}
    \label{tab:bd-rates-complexity}
\end{table}

\Cref{tab:bd-rates-complexity} compares the BD-rates of the proposed HyperCool against HEVC, along with encoding complexity.
It shows that HyperCool improves compression over N-O Cool-chic, with only a slight increase in encoding cost.
We also compare HyperCool’s encoding complexity to standard Cool-chic using the \textit{fast} and \textit{slow} presets from the official open-source implementation \cite{coolchic}.
HyperCool is 500 to 3000 times cheaper to encode than fully overfitted Cool-chic, though at the cost of reduced compression performance.

\subsection{Modulations rate overhead and usage\label{subsec:switch}}

Adapting decoder parameters to the image using modulation parameters $\Delta_{\mathbf{w}}$ requires transmitting them, adding rate overhead.
Therefore, modulations are only used if the compression improvement outweighs their signaling cost.
This is determined at the encoder via a simple test, which disables modulations when counterproductive.

% \newline

\begin{figure}[!t]
\centering
\begin{tikzpicture}
\begin{axis}[
    width=\linewidth,
    height=7cm,
    xlabel={Average compressed image size [MByte]},
    ylabel={Modulation usage [\%]},
    ymin=0, ymax=100,
    xmin=0, xmax=10,
    xtick distance=2,
    ytick distance=20,
    grid=both,
    legend style={at={(0.9,0.1)}, 
        column sep=1.5em, anchor=south east, legend columns=1},
    every node near coord/.append style={font=\footnotesize},
    mark options={solid},
    thick,
]
% Kodak: 9437184 pixels
\addplot[mark=*, color=myblue, thick] table [x=MBytes, y=usage_modulation_percents] {data/modulation_usage/kodak.tsv};
% CLIC2020-pro-valid: 91451931 pixels
\addplot[mark=square*, color=myred, thick] table [x=MBytes, y=usage_modulation_percents] {data/modulation_usage/clic20-pro-valid.tsv};
\legend{Kodak, CLIC2020}
\end{axis}
\end{tikzpicture}
\caption{Usage of the modulation parameters $\Delta_{\mathbf{w}}$ across different bitrates.}
\label{fig:modulation-percent-line}
\end{figure}
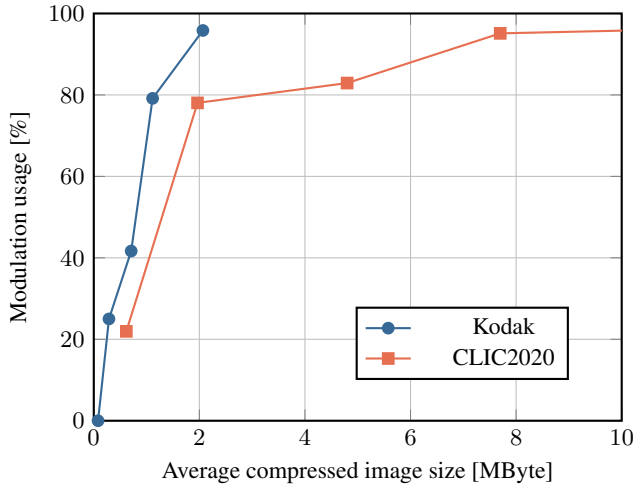

\Cref{fig:modulation-percent-line} shows the proportion of images using modulations under different rate constraints and datasets.
At higher rates, nearly all images use the hypernetwork modulations, as more bits are available for parameter signaling.
However, under stricter rate constraints, many images do not use modulations \textit{e.g.}, only 20~\% of the images at the lowest rate on CLIC2020.
This behavior explains the improved performance of HyperCool on CLIC2020, where larger images permit greater use of modulations due to higher bit budgets.
% \newline

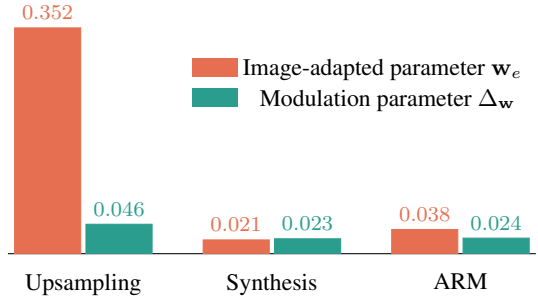
\begin{figure}[!t]
  \centering
  \begin{tikzpicture}
    \begin{axis}[
      ybar,
      bar width=25pt,
      width=\linewidth,
      height=5cm,
      symbolic x coords={Upsampling, Synthesis, ARM},
      hide y axis,
      axis x line*=bottom,
      xtick=data,
      ymin=0, ymax=0.4,
      enlarge x limits=0.2,
      major x tick style = transparent,
      nodes near coords,
      nodes near coords align={vertical},
        every node near coord/.append style = {
            anchor=south,
            font=\footnotesize,
            /pgf/number format/.cd,
            /pgf/number format/fixed,
            /pgf/number format/fixed zerofill,
            /pgf/number format/precision=3,
        },
        legend style={
        at={(1,0.8)},
        anchor=north east,
        legend columns=1,
        draw=none,
        fill=none,
        font=\small,
      },
      legend image code/.code={
         \draw[fill=#1, draw=none] (0cm,-0.1cm) rectangle (0.6cm,0.15cm);
      },
    ]
    \addplot+[myred] coordinates {
      (Upsampling,0.35151628)
      (Synthesis,0.02144581)
      (ARM,0.037649225)
    };
    \addplot+[mygreen] coordinates {
      (Upsampling,0.045820083)
      (Synthesis,0.023213154)
      (ARM,0.024242828)
    };
    \legend{Image-adapted parameter $\mathbf{w}_e$, \hphantom{a}Modulation parameter $\Delta_{\mathbf{w}}$}
    \end{axis}
  \end{tikzpicture}
  \caption{Comparison of the standard deviation of image-adapted and modulation parameters on the CLIC2020 dataset.}
  \label{fig:std_lambda_distributions}
\end{figure}

\Cref{fig:std_lambda_distributions} illustrates that modulation parameters $\Delta_{\mathbf{w}}$ are more compact than the full parameters $\mathbf{w}_e$.
We confirm this by comparing the standard deviations of $\Delta{\mathbf{w}}$ and $\mathbf{w}_e$, computed from a Laplace distribution fitted to the parameters.
Modulations show lower variance, indicating better compressibility with Exp-Golomb coding.

\subsection{Hypernetwork ablation experiments\label{subsec:ablation}}

To assess the contribution of each hypernetwork module, we start from the full HyperCool model and selectively disable modulations for different components.
\Cref{tab:ablation} summarizes the average change in bitrate and PSNR compared to the base N-O Cool-chic across different rate points.
Using only ARM modulations reduces the latent bitrate without improving PSNR.
In contrast, applying only upsampling and synthesis modulations improves PSNR but increases the total bitrate due to the modulation overhead.
Combining all modulations balances these effects, yielding a total bitrate reduction of 0.011 bpp and a PSNR increase of 0.071 dB.
These results highlight the complementary nature of the modulation modules: ARM modulations reduce latent rate, while upsampling and synthesis modulations enhance reconstruction quality.
Together, they achieve gains in both compression rate and image quality, albeit with a slight increase in modulation bitrate.

\definecolor{okgreen}{HTML}{0F9D58}
\definecolor{kored}{HTML}{DB4437}

\newcommand{\greencheck}{{\color{okgreen}\ding{51}}}
\newcommand{\ingreen}[1]{\textcolor{okgreen}{#1}}
\newcommand{\inred}[1]{\textcolor{kored}{#1}}

\begin{table}[!t]
    \small
    \centering
    \caption{Change in rate and PSNR compared to N-O Cool-chic when using different modulations. Averaged across rates.}
    \begin{tblr}{
        colspec={Q[0.5cm,c] Q[0.4cm,c] Q[0.4cm,c] Q[1.3cm,c] Q[0.9cm,c] Q[0.9cm,c] Q[1cm,c]},
        row{1} = {darkgray!5, font=\bfseries},
        row{2} = {darkgray!5},
        cell{1}{1}={c=3,r=1}{},
        cell{1}{4}={c=3,r=1}{},
        cell{1}{7}={c=1,r=2}{},
        vline{4}={1-Z}{solid, fg=darkgray,1pt},
        vline{7}={1-Z}{solid, fg=darkgray,1pt},
    }

        Modulations & & & Rate [bpp] $\downarrow$  & & & PSNR [dB] $\uparrow$ \\
        ARM & Ups & Syn      &   Modulation & Latent & Total    &   \\
        \cmidrule[1pt,darkgray]{1-Z}
        \greencheck & \greencheck & \greencheck & +0.008 & -0.019 & \ingreen{-0.011} & \ingreen{+0.071} \\
        \greencheck &             &             & +0.003 & -0.019 & \ingreen{-0.016} & 0 \\
                    & \greencheck & \greencheck & +0.005 & 0 & \inred{+0.005} & \ingreen{+0.071} \\
    \end{tblr}
    \label{tab:ablation}
\end{table}

\subsection{HyperCool as an overfitting initialization \label{subsec:samerd}}

Standard Cool-chic encodes an image through the overfitting of the latent representation and decoder parameters, starting from a random initialization.
Both N-O Cool-chic and the proposed HyperCool provide a strong initial guess for the latent and decoder parameters, improving initialization for subsequent overfitting.
% \newline

\Cref{fig:bd-vs-flops} compares Cool-chic encoding using three different initializations: random, N-O Cool-chic, and HyperCool.
Across all encoding complexities, HyperCool initialization consistently outperforms N-O Cool-chic, highlighting the hypernetwork’s effectiveness.
Moreover, HyperCool enables reaching HEVC-level compression 40\% faster than random initialization.
However, standard Cool-chic with random initialization achieves better asymptotic performance, suggesting HyperCool may converge to a local minimum.

\section{Limitations and Future Directions}

Although our results are positive, there are notable limitations that must be further investigated.
The performance advantage of our hypernetwork is most pronounced at medium to high bitrates.
At low bitrates, the quantization process often favors excluding the hypernetwork's output to save on the additional rate, leading to performance nearly identical to the underlying N-O Cool-chic model.
Additionally, our approach depends on the quality of the pre-trained N-O Cool-chic base model, as the hypernetwork only generates modulation parameters for it.
% \newline

Future work could explore several promising directions.
Alternative hypernetwork architectures may yield further improvements.
Furthermore, it would be valuable to contrast the HyperCool approach with other meta-learning strategies.
For example, COIN++ \cite{dupont_coin_2022} and MLIIC \cite{zhang_mliic_2025} successfully apply MAML \cite{finn_model-agnostic_2017} to learn a base network serving as a starting point for task-wise adaptation.
A hybrid method combining MAML-based adaptable bases with our hypernetwork modulation could better parametrize the base model, improving BD-rate while keeping computational cost unchanged.

\section{Conclusion}

This work addresses the main drawback of overfitted image codecs: their slow and computationally expensive encoding process that requires per-image optimization.
We introduced a novel hypernetwork architecture that builds upon the Non-Overfitted Cool-chic framework to generate image-adaptive parameters in a single forward pass.
This approach improves compression efficiency without per-image optimization, providing a step toward practical use of overfitted codecs.
% \newline

Our method achieves a 4.9\% BD-rate reduction over the N-O Cool-chic baseline with minimal computational overhead.
% The cost is under two forward passes of the N-O Cool-chic analysis transform, delivering competitive rate-distortion performance at a low computational budget.
Additionally, the hypernetwork output provides a strong initialization for full Cool-chic decoder optimization, reducing the number of fine-tuning steps by 40\%.
This makes our approach a practical way to accelerate overfitted codecs and broaden their range of applications.

 \bibliographystyle{IEEEbib}
% \bibliography{strings,refs}
\bibliography{refs.bib}

\end{document}